\documentclass{aa} 
\usepackage{epsfig}
\usepackage{rotating}
\usepackage{txfonts}
\usepackage{aalongtable,lscape}

\usepackage{txfonts}

\newcommand{\teff}{T$_{\rm eff}$}

\newcommand{\feh}{$[\rm Fe/\rm H]$}

\begin{document}
  \title{Metallicity of low-mass stars in
    Orion 
  \thanks{Based on observations collected at Paranal Observatory,
ESO (Chile). Programs 072.D-0019, 074.C-0757, 076.D-0136,
076.C-0145.}}
          
  \subtitle{}

  \author{V. D'Orazi\inst{1,2} \and S. Randich\inst{2} 
         \and E. Flaccomio\inst{3} \and F. Palla\inst{2} 
           \and G.G. Sacco\inst{3} \and
            R. Pallavicini\inst{3}\thanks{
We dedicate this paper to
the memory of Roberto Pallavicini who prematurely passed away during the
completion of this work. We all miss him sorely.}
          }

  \offprints{V. D'Orazi, email: vdorazi@arcetri.astro.it}

  \institute{Dipartimento di Astronomia e Scienza dello Spazio,
Universit\`a di Firenze, Largo E. Fermi 2, I-50125 Firenze, Italy
\and
INAF-Osservatorio Astrofisico di Arcetri, Largo E. Fermi 5,
I-50125 Firenze, Italy
\and
INAF-Osservatorio Astronomico di Palermo, Piazza del
Parlamento 1, I-90134 Palermo, Italy
}

\titlerunning{}
\date{Received Date: Accepted Date}


\abstract
{Determining the metal content of low-mass members 
of young associations provides a tool that addresses different issues,
such as triggered star formation or the link between the metal-rich
nature of planet-host stars and the early phases of planet formation.
The Orion complex is a well known 
example of possible triggered star formation
and is known to host a rich variety of proto-planetary disks around its
low-mass stars. Available metallicity measurements yield discrepant 
results.}
{
We analyzed FLAMES/UVES and Giraffe spectra of low-mass members
of three groups/clusters belonging to the Orion association.
Our goal is the homogeneous determination of the metallicity 
of the sample stars, which allows us
to look for [Fe/H] differences between the three regions
and for the possible presence of metal-rich stars. 
}
{Nine members of the ONC and one star each 
in the $\lambda$~Ori cluster and OB1b subgroup were analyzed. 
After the veiling determination, we retrieved the metallicity by
means of equivalent widths and/or spectral
synthesis using MOOG.}
{We obtain an average metallicity for the ONC
$\feh~=-0.01\pm 0.04$. No 
metal-rich stars were detected and the dispersion within our sample is 
consistent with measurement uncertainties.
The metallicity of the $\lambda$~Ori member is also solar, while
the OB1b star has an [Fe/H] significantly below the ONC average. If confirmed by
additional [Fe/H] determinations in the OB1b subgroup,
this result would support the triggered star formation 
and the self-enrichment scenario for the Orion complex. 
}
{}

   \authorrunning{V. D'Orazi et al.}
   \maketitle
\section{Introduction}\label{intro}
The Orion OB1 association, located at roughly 400~pc,
is probably the best known region of active star formation.
It comprises four major subgroups
of different ages, sizes, and positions (Ia, Ib, Ic, Id where the Orion
nebula cluster -ONC- is), together with several grouping of stars
and clusters, covering a wide range of ages and environmental
conditions: extremely young groups of stars are present still embedded in their
parental clouds, 3--5 Myr old clusters like
$\sigma$ and $\lambda$ Ori, as well as older populations
like the Ori Ia sub-association (age $\sim 10$~Myr).
Orion is since long considered as a possible example of
triggered star formation (Blaauw 1991, ASI Ser. 342, p.~125), supported by
the increasing ages between the Id/Ic, Ib, and Ia subgroups and by the
different content of gas and dust of the different regions.
In the triggered scenario,
winds and SN-driven shock waves
originating from a first generation of massive stars induce the formation
of a new group/cluster of stars. This second generation population is likely
contaminated by the
enriched ejecta of the massive stars of the first generation
(e.g., Reeves \cite{reeves}) and thus might show a different abundance
pattern. The secure and homogeneous
determination of the chemical composition across Orion,
and the comparison of the abundance patterns of the different subgroups, 
hence represent a critical and independent test of this scenario.

Besides providing insights into triggered star formation models, the 
determination of the metallicity of low-mass stars in young associations
is important in the context of extra solar planets.
It is now well-established that 
planet-host stars are on average more metal-rich than
stars without planetary-mass companions (Gonzalez~\cite{gon97};
Gonzalez et al.~\cite{gon2001}; 
Santos et al.~\cite{santos01};
Fisher \& Valenti~\cite{fis05}).
The mean metallicity of stars hosting a gas giant 
planet is [Fe/H]$= +0.15 \pm 0.23$, compared to the value 
[Fe/H]$=-0.1\pm 0.18$ for the solar neighborhood.
Also, the frequency of giant planets around metal-rich 
stars ([Fe/H]$\geq 0.3$) is f$_{\rm p}$ $\geq$ 20\%, much higher
than that of planets around solar (or under-solar) metallicity stars
(f$_{\rm p}$ $\sim$ 3\%). More important, complementary studies 
suggest that the high metal content is primordial and not
due to pollution by cannibalized planetary bodies
(Ecuvillon et al.~\cite{ecuvillon}; Gilli et al.~\cite{gilli},
Pinsonneault et al.~\cite{pins01};
see however Pasquini et al.~\cite{pas07} for
a different view). 

Surveys of the properties of planet-host stars 
provide clues to the mechanism of planet formation
based on its final product (i.e., the already formed planet), 
since the majority of
field stars where a giant planet has been detected are older than
$\sim 1$~Gyr. On the other hand,
circumstellar disks around pre-main sequence (PMS) stars, which 
are the birth sites of planets, are ubiquitously found in star-forming regions
(SFRs). The question then is what the metallicity of low-mass
stars that are likely to be forming planets now in SFRs and young associations
might be. To definitively answer it, the metallicity needs to 
be derived, as much as possible, for a 
complete SFR sample. The Orion complex already represents
an excellent target,
given the high frequency and wide variety of proto-planetary disks
around its low-mass members.

In the last 50 years,
hundreds of studies have investigated key aspects
related to the Orion stellar population, as well as to the associated
HII regions and the surrounding
interstellar medium. However, considerably less attention has been
given to determining its chemical composition. 
Available information on the abundance pattern in Orion and, in particular, of
the ONC, 
mostly comes from studies of the ionized gas in the Orion nebula or
from abundance determination in stars of early spectral-type. 
The latter, however,
is affected by several uncertainties, for example, NLTE and rotational
broadening of the spectral lines.
Few measurements of the metallicity among lower-mass stars
have so far been carried out, as we discus in Sect.~\ref{status} below. 
With this background in mind, we exploited the available
high-resolution spectra acquired in the framework of two different projects 
(Palla et al. \cite{pal05},~\cite{pal07}; Sacco et al.
\cite{sacco08}) to measure the metallicity in a sample of ONC members
and in one star
belonging to the $\lambda$~Ori cluster. One member
of the Orion OB1b subgroup, whose
spectrum was retrieved from the ESO archive, was also analyzed.

The paper is organized as follows. In Sect.~2 we give an overview of
previous determination of metallicity in Orion members; in Sect.~3
we describe the sample, 
observations, and data reduction. The analysis is presented in Sect.~4 and
the results and discussion in Sect.~5, followed by the conclusions
(Sect.~6).
\section{Previous metallicity determinations in Orion}\label{status}
Based on the determination of oxygen in the
Orion nebula by Osterbrock et al. (\cite{oster}),
it has been believed until recently that the metallicity of Orion
is lower than that of the Sun, representing a problem
for Galactic evolution models in the solar vicinity.
While the reappraisal of oxygen in the Orion {\it nebula}
by the study of Esteban~(\cite{esteban}) no longer supported this view,
{\it stellar} abundances obtained
by Cunha \& Lambert (\cite{cun94}) and Cunha et al. (\cite{cun98})
also suggested a lower-than-solar metallicity for the whole
Orion complex. Specifically,
Cunha and collaborators measured 
Fe, O, C, N, and Si
abundances in a sample of B- and F/G- type Orion members, primarily belonging
to the subgroups 1a and 1c.
They found that the metallicity ([Fe/H]) 
ranges between $-0.31$ (Ic, G-type star) and $+0.14$ (Ia, B-type star); 
in spite of this rather wide interval Cunha and collaborators concluded that
all the four subgroups 1a, 1b, 1c, and 1d (where the ONC
is located) are characterized
by the same, somewhat below solar, metallicity
([Fe/H]$=-0.08 \pm 0.13$, where the error
is the 1$\sigma$~deviation from the mean). 
F and G-type stars only would indicate an even lower metal content
([Fe/H]$=-0.13\pm 0.13$). Interestingly, the most metal-rich
among the F/G-type stars is the only ONC members
included in the sample. For this star, P1455, they derived a metallicity
[Fe/H]$=+0.08 \pm 0.15$. At variance with their claim of a homogeneous iron
content,
Cunha and collaborators (\cite{cun92},~\cite{cun98}) 
found evidence of star-to-star variations in O and Si abundances,
with a few O- and Si-enhanced stars being more
centrally located (in the Trapezium region) than the
O- and Si-poor ones, distributed throughout the association.
The authors suggest that the observed abundance pattern
could be evidence of the expected self-enrichment.
More recently, Sim\'on-D\'\i az et al.
(\cite{sim06}) derived oxygen abundances for a few
B-type stars in the Trapezium
finding lower values than those of Cunha \& Lambert~
and, in contrast to their results, no evidence of any elemental enhancement.

Focusing on low-mass stars,
seven objects in Orion 1c and 1d were included in the study
of Padgett (\cite{padgett}), who derived [Fe/H] values in the interval
$-0.1$ to +0.23~dex and an average
[Fe/H] slightly above solar.
Of the four ONC stars included in the sample, two showed
slightly over-solar abundances ([Fe/H]$=0.14\pm 0.18$ and $0.08\pm 0.17$
respectively), and the other two stars
a solar metallicity ([Fe/H]$=-0.01 \pm 0.21, -0.01\pm 0.17$).
Finally, 
based on three stars, Santos et al. (2008) obtained an average for the
ONC [Fe/H]$=-0.13 \pm 0.06$, again below solar.
In summary, existing determinations
of the metal content in the whole Orion complex
reach controversial conclusions both on the overall metallicity
--in particular on  whether it is below solar or not-- on
the presence of a star-to-star scatter within a given subgroup
(the ONC in particular), and on differences in [Fe/H] between
the different subgroups.
\section{Sample stars and observations}
\subsection{ONC}
The spectra were acquired in two observing sessions with
FLAMES on VLT/UT2 (Pasquini et al. 2000)
as part of a project focused on measurements
of lithium abundances in very low-mass members of the ONC
(Palla et al.~\cite{pal05},~\cite{pal07}). 
In both runs Giraffe
fibers were mostly allocated to M-type stars; instead, UVES fibers were 
used to observe K- and M-type stars with the goal of deriving
their accretion rates by means of the Ca II infrared triplet 
(Flaccomio et al., in prep).
The present sample includes all the K-type stars
observed with UVES (seven stars), while M dwarfs were discarded 
since determination of
metallicity in these cool stars is not feasible with standard methods/codes
and Kurucz model atmospheres
(see Sect.~\ref{analysis} below). 
To the UVES sample, we added the only two K-type stars observed
with Giraffe and for which Palla et al. (\cite{pal05}) derived a negligible
veiling. The total sample hence consists of nine stars.

The observations were carried out on 15 February (Visitor Mode)
and 15 November (Service Mode), 2004. The
UVES spectrograph (Dekker et al. 2000) in fiber mode
provides a resolution R=40000. 
We used the CD4 cross-disperser (needed to target the Ca II triplet),
allowing us to
cover a spectral range from $\sim$6700~\AA~to 10000~\AA.
Giraffe was used with the HR15 setup, yielding a resolution R$\sim$ 19000
and spectral coverage from 660.7 to 696.5 nm.\\
The sample stars are listed in Table~\ref{onc_targets}, where we
give the identifier from Hillenbrand (\cite{hil97}), an observing 
log (the dates of 
the observations, total exposure time, S/N ratios), along with
the stellar properties
(I magnitude, V--I color, spectral-type, rotational period,
effective temperature --T$_{\rm eff}$--, and K-band excess). 
are also listed in the table.
Information on photometry, spectral-types, effective temperatures, and excess
were taken from Hillenbrand (\cite{hil97}) and Hillenbrand et al. 
(\cite{hil98}), while rotational periods come from 
Herbst et al.~(\cite{her02}). 

\begin{figure*}
\includegraphics[angle=270, width=16cm]{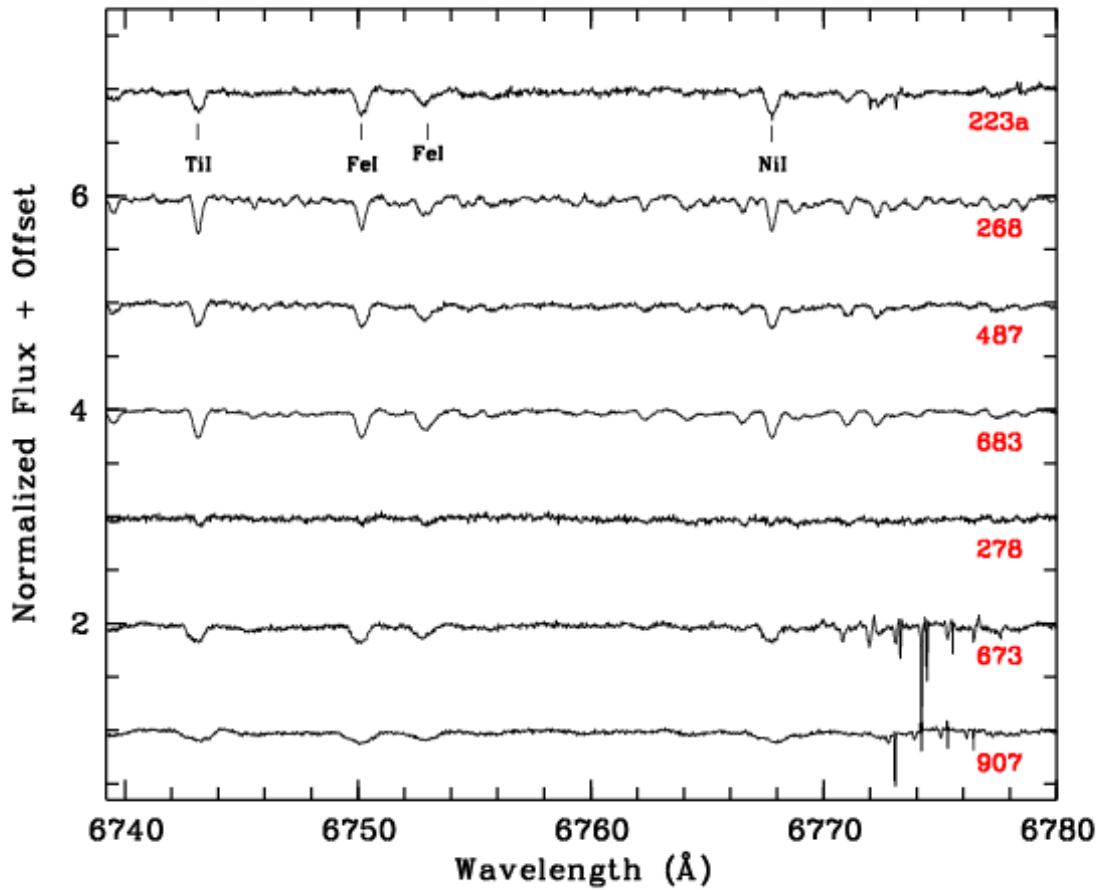}
\caption{UVES 
spectra of the ONC sample stars in the wavelength range 6740--6780
 \AA. Different features are marked. Note the shallower lines 
in the spectrum of star h278, for which we derived a high veiling value.}
\label{onc_spectra}
\end{figure*}

Reduction of UVES data was carried out using 
the FLAMES-UVES pipeline, 
following the standard procedure: bias subtraction, flat-field correction, 
order extraction and wavelength calibration. Sky subtraction was performed
outside the pipeline context, using the fiber allocated to the sky.
Spectra of the same star obtained in
different exposures were then co-added, after checking that no
radial velocity variations were present.
Giraffe data were reduced using the BLDRS software
(Blecha \& Simond \cite{blecha} -see details in Palla et al.~\cite{pal07}).
Final S/N ratios are in the range 70--125 for the UVES spectra and 150--200
for the Giraffe ones. All
UVES spectra are shown in Fig.~\ref{onc_spectra}.
\begin{figure*}
\includegraphics[angle=-90, width=16cm]{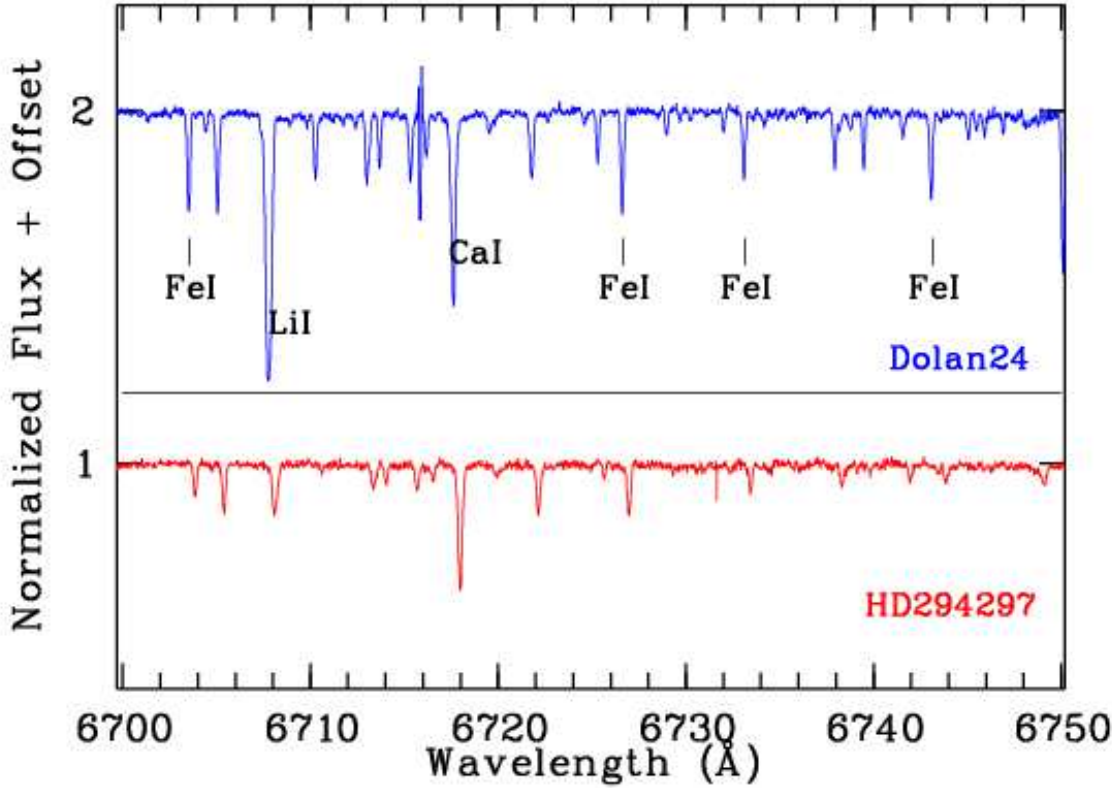}
\caption{Same as Fig.~\ref{onc_spectra}, but UVES spectra of the $\lambda$
Ori star (upper panel) and of the OB1b member
are shown. Note the different spectral
interval displayed here with respect to Fig.~\ref{onc_spectra}.}
\label{sigma_spectra}
\end{figure*}
\setcounter{table}{0}
\begin{table*}	
\caption{ONC target stars, observing log, and target properties$^1$.
}			
\begin{tabular} {rcccccclcccc}
\hline
Star & Obs. Dates & Instrument & Exp. time & S/N & I  & V--I &  
Sp. Type &  T$_{\rm eff}$ & $\Delta$(I-K) & P & v$_{\rm rad.}$\\
(H97)&            &   & (sec)    &     &      &    &  & (K) & (mag) & (days) &
(km/s)\\
(1) & (2) & (3) & (4) & (5) & (6) & (7) & (8) & (9) & (10) & (11) & (12)\\
 \hline
 223a &  15-02-04 & UVES, CD4 & 3$\times$3600  & 70   & 13.64 & 2.42 & K5   & 4395 & 0.48 & -- & $25.43\pm 1.58$\\
 268 &  15-02-04 & UVES, CD4 &  3$\times$3600  & 95   & 12.88 & 1.61 & K5-K6  & 4197 & 0.16 & 9.81 & $27.80 \pm 1.02$\\ 
 278 &  15-02-04 & UVES, CD4 &  3$\times$3600  & 70   & 13.66 & 1.94 &  K2-K7e  & 4395 & 1.69 & 6.76 & $24.58 \pm 1.49$\\
 487 &  15-02-04 & UVES, CD4 &  3$\times$3600  & 105   & 12.79 & 1.78 & K6     & 4197 & 0.28 & -- & $29.27 \pm 0.87$\\
 673 &  15-02-04 & UVES, CD4 &  3$\times$3600  & 110   & 13.07 & 1.83 & K5 & 4395 & 0.81 & 1.44 &  $24.53 \pm 1.29$\\
 683 &  15-02-04 & UVES, CD4 &  3$\times$3600  & 130   & 11.76 & 1.85 & K6    & 4197 & 0.28 & 11.50 &  $25.73\pm 1.51$\\
 907 &  15-11-04 & UVES, CD4 &  4$\times$2600  & 125   & 12.77 & 1.60 & K1-K4  & 4775 & --& -- &  $20.82\pm 1.24$\\ \hline
 664 &  15-02-04  & GIRAFFE & 3$\times$3600  & 150   & 14.05 & 2.88 & K5.5-K7 & 4197 & $-0.18$ & 7.20 & $25.36 \pm 1.18$\\
1020 &  15-02-04  & GIRAFFE & 3$\times$3600  & 200   & 12.65 & 1.40 & K6     & 4197  & $-0.01$ & -- & $27.60 \pm 1.82$\\
 \hline 
\end{tabular}\\
\footnotemark[1]{
Identifier~(1),
I, V--I, spectral-types (6, 7, 8), and T$_{\rm eff}$ values (9
--from spectral-types using the scale of Kenyon \& Hartmann \cite{kh95}) 
come from Hillenbrand (\cite{hil97}); 
K-band excess (10) and rotational periods~(11) were taken from 
Hillenbrand et al. (\cite{hil98}) and Herbst et al.~(\cite{her02}),
respectively. Radial velocities
(12) were measured from our spectra.}
\label{onc_targets}
\end{table*}
\setcounter{table}{1}
\begin{table*}
\caption{$\lambda$~Ori and OB1b targets, observing log, and target properties$^1$.
}			
\begin{tabular}{clcccccccc}
\hline
Star & Obs. Dates & Instrument & Exp. time & S/N & I  & R--I & Sp.Type & T$_{\rm eff}$ & v$_{\rm rad}$\\
&            &  & (sec)    &     &    &      &        & (K) & (km/s)\\
(1) & (2) & (3) & (4) & (5) & (6) & (7) & (8) & (9) & (10) \\ \hline
HD~294297  & 2005-11-17     &UVES, CD4 &  3x3575  &  150     &  ---       &  ---      &  G0 & 6100  & $26.46 \pm 0.41$ \\
  & & & & & & & &  & \\ 
Dolan24  & 2005-10-16/17/18  &UVES, CD3 &  4x2760     &  200         & 11.75     &  0.53     &   ---  & 5080 & $22.83\pm 0.53$\\ 
         & 2005-11-11     &   &       &       &           &     &     &  \\
\hline  
\end{tabular} \\
\footnotemark{
 Information for star HD~294297
comes from the SIMBAD database. Identifiers and photometry of Dolan 24 
in $\lambda$~Ori were taken from Dolan \&
Mathieu (\cite{dm99}). \teff~for HD~294297 was retrieved from Cunha et al.
(\cite{cun98}), while that of Dolan 24 was estimated from the R--I color.
Radial velocities were measured from our spectra.}
\label{sigma_targets}
\end{table*}
\subsection{$\lambda$~Ori and Orion OB1b}
As in the case of ONC, we selected stars of 
$\lambda$ Ori from a wider sample acquired to investigate 
membership and accretion properties
in this young cluster by Sacco et al.~(\cite{sacco08}).
To select the stars, we applied the same criteria used for the ONC sample,
discarding the fast rotating members (v$\sin i \ge 30$~km/s), binaries, and
stars with spectral-type later than K7; this criterion
resulted in the selection of one star only (Dolan 24). 
The observations were carried out in Service Mode during October and 
November, 2005, using  FLAMES-UVES 
and CD\#3 cross-disperser, covering the spectral interval $\sim 4840-6840$~\AA.
We enlarged our sample with a member of the OB1b association
(HD~294297) whose FLAMES-UVES spectrum, obtained with the 
CD\#4 cross-disperser, was retrieved from the ESO Archive 
(Program 076.C-0145, PI Jeffries). This star is included in the sample
of Cunha et al. (\cite{cun98}) and thus allows a direct comparison of their and
our abundance scales.
For both stars data reduction was 
performed under ESO-MIDAS context and using the FLAMES/UVES pipeline. 
The two targets along with information on their properties
and observations are listed in
Table~\ref{sigma_targets}; their spectra are shown in Fig.~\ref{sigma_spectra}.
\section {Analysis and abundance determination}
\subsection {Radial velocities}
Radial velocities were computed from our
spectra using the task RVIDLINES in IRAF\footnote{
IRAF is distributed by the National Optical
Astronomical Observatories, which  are operated by the Association of
Universities for Research in Astronomy,  under contract with the National
Science Foundation.} and a list of 10 lines; with the exception of Dolan~24,
all measurements are
in good agreement with previous determinations in the literature
(Sicilia-Aguilar et al.~\cite{sicilia};  Caballero
~\cite{cab07}). For Dolan~24 we instead obtain a radial velocity
lower than the measurement of Dolan \& Mathieu (\cite{dm99} --29.14~km/s) and
slightly below two different estimates of the
average radial velocity of $\lambda$~Ori (v$_{\rm rad}=24.3\pm 2.8$~km/s
--Dolan \& Mathieu~\cite{dm99}; v$_{\rm rad}=27.3\pm
0.49$~km/s --Sacco et al.~\cite{sacco08}), suggesting that this star
likely is a spectroscopic binary. Derived radial velocities for all stars
are listed in the last columns of Tables~\ref{onc_targets} 
and \ref{sigma_targets}
\subsection {Estimate of veiling}
As well known, spectra of PMS stars might be 
affected by spectral veiling that alters measured
equivalent widths (EWs), since 
photospheric absorption lines are filled in by accretion shock emission. 
As a consequence, the observed lines are weaker than
intrinsic ones and one needs to correct for veiling before deriving
abundances. This is accomplished by 
determining $r$, the ratio of the excess to the photospheric continuum;
then the relationship between the true and measured EWs is
EW$_0$ = EW$_{\rm meas.}\times(1+r)$. 

As mentioned in the previous section, veiling for the two ONC
stars observed with
Giraffe was already derived by Palla et al.~(\cite{pal05}); both of them
have $r \approx 0$, and this was indeed the reason why we included them
in our sample. On the other hand,
we directly estimated $r$ from our spectra
for the seven ONC stars observed with UVES
and for the $\lambda$~Ori member (HD~294297 should not be affected
by veiling --Cunha et al.~\cite{cunha95}),
following a procedure similar to that described in Palla et al.
(\cite{pal05},~\cite{pal07}), 
but using different features. Namely,
we measured the EWs of selected strong lines
in the target stars and compared
them with those measured in the spectra of 11 members
of the IC~2602
and IC~2391 clusters, which are old enough ($\sim$30--50~Myr) to ensure that
their spectra are not affected by veiling.
The IC clusters stars cover the same range of effective temperatures as
the Orion targets and their spectra are characterized by
a similar resolution 
(see Stauffer et al.~\cite{stauffer};
Randich et al.~\cite{R01}). 
For the 580~set-up we used nine lines;
namely, Ca~{\sc i} 5857.5~\AA, Ca~{\sc i} 6102.7~\AA, 
Ca~{\sc i} 6122.2~\AA, Fe~{\sc i} 6546.3~\AA, 
Ni~{\sc i}~6643.6~\AA,
Fe~{\sc i} 6662.5~AA, V~{\sc i}~6624.8~\AA, 
Ca~{\sc i} 6717.7~\AA, Ti~{\sc i} 6743.2~\AA.
For the 860~nm set-up we 
employed six lines in the range 6743--8085 $\AA$; 
namely, 6743.12 $\AA$ (Fe~{\sc i}), 7148.15 $\AA$ (Ca~{\sc i}), 
7445.75 $\AA$ (Fe~{\sc i}), 
7698.97 $\AA$ (K~{\sc i}), 7937.13 $\AA$ (Fe~{\sc i}), 
7998.97 $\AA$ (Fe~{\sc i}).

As a first step, we checked whether a dependence of the strength of
the lines on
temperature was present among IC~cluster members:
with the exception of the Ca~{\sc i} 6122.2~\AA~and K~I 7698.97~\AA~features, 
all the other lines did not show strong trends with T$_{\rm eff}$ within
$\sim 300$~K. 
Hence, in these cases we derived the mean EW considering
all stars in the two clusters and used this average value for 
comparison with the Orion stars. 
For the two features whose strength depends on effective temperature,
we instead used the EW value of the IC cluster star with T$_{\rm eff}$ 
closest to that of the given ONC star.
For each line, the quantity $r_{\rm line}$ was calculated as
EW$_{\rm IC}/$EW$_{\rm ONC}-1$. We did not find any trend of $r_{\rm line}$
with wavelength for none of the stars, similarly to the results of Santos
et al.~(\cite{santos08}). Hence, we computed the average value of 
$r$ for each star using all lines. These final values are listed in Col.~2 of 
Tables~\ref{onc_parameters}  and~\ref{sigma_parameters}. 
The two tables show that $r$ is consistent
with zero for all stars with exception of the ONC members h278, for which
we obtained $r=0.8 \pm 0.07$; due to this relatively
high veiling, the star was discarded from the sample.
\subsection{Abundance analysis}
The analysis was performed using MOOG (Sneden~\cite{sne73} 
--2002 version)
and a grid of 1-D model atmospheres from Kurucz (\cite{kur93}), 
by means of both
EWs and spectral synthesis. 
Radiative and Stark broadening are treated in a standard way in MOOG;
as for collisional broadening, we used the Uns\"old approximation (\cite{uns})
for all the lines. As discussed by Paulson et al. (\cite{paul03}) this choice
should not greatly affect the differential analysis with respect
to the Sun.  We also mention that
very strong lines (EW$\geq 150$~m\AA)
that are most affected by the treatment of damping
have been excluded from our analysis.
\subsubsection{EW analysis and stellar parameters} \label{analysis}
As mentioned earlier, the spectrum of the 
$\lambda$~Ori star was acquired using the CD3 and the analysis was done 
using the line list of Randich et al.~(\cite{randich_m67}) who
had used the same UVES set-up. For the ONC targets
and the spectrum retrieved from the archive that were observed with the CD4,
we instead built a new line list. Although the spectra cover
the wavelength interval 6700 to 10000~\AA, only the blue part
up to $\sim 8000$~\AA~was usable, since the red part
was contaminated by several telluric lines which could not be corrected,
owing to the lack of spectra of early-type stars.
Fe~{\sc i} features to be used for the analysis in the interval
$\sim 6700-8000$~\AA~were selected from different 
sources in the literature
and subsequently checked for suitability
(e.g. for blends) both on the solar spectrum observed with UVES and on
the best quality sample spectra.
The final list contains 40 Fe~{\sc i} lines for the warm star
HD~294297 and 26 lines for
the cool members of the ONC (see Table~\ref{sun}); 24 lines are in common.
Not for all the cool stars, however, it was possible to measure
all the 26 lines. Most
$\log~gf$ values were taken from Clementini et al.~(\cite{clementini});
for the few lines not included in that study, $\log~gf$ were instead
retrieved from the Vienna Astronomical Line Database (VALD\footnote
{http://ams.astro.univie.ac.at/cgi-bin/vald/}).
We performed the analysis of the solar spectrum obtained
with UVES at the same resolution as that of our sample spectra
using both the line list with 40 lines and that with 26
lines. Adopting as solar
parameters T$_{\rm eff\;\odot}=5770$~K, $\log g_{\odot}=4.44$
and $\xi_{\odot}=1.1$~km/s (see Randich et al.~\cite{randich_m67}),
with both line lists 
we obtained $\log$~n(Fe)=7.50$\pm 0.03$, a value very close to
$\log$~n(Fe)$_\odot$=7.52 (Anders \& Grevesse,~\cite{ag89}).
Individual $\log$n(Fe) values for the Sun are listed in Table~\ref{sun}.
[Fe/H] for the sample stars with CD4 spectra
was derived differentially with respect to our own
determination. For Dolan~24, analyzed using the line list of Randich
et al., a value $\log$n~(Fe)$_{\odot}=7.52$ was instead adopted.

EWs were measured using the package
SPECTRE and a gaussian fitting procedure.
Although the spectra had been previously normalized, local continuum
was inspected and, if needed, adjusted at each EW measurement.
Initial stellar parameters were estimated as follows:
{\it i.} effective temperatures for the ONC stars were 
retrieved from Hillenbrand~(\cite{hil97}) who, in turn, had derived
them from spectral-types using the scale of Kenyon \& Hartmann~(\cite{kh95}). 
T$_{\rm eff}$
of HD~294297 was taken from Cunha et al. (\cite{cun98}), while
\teff~for the $\lambda$~Ori star, for which the spectral-type
is not available, was derived
from the R--I color and again the scale of Kenyon \& Hartmann.
A reddening E(R-I)=0.07 was adopted (see discussion in
Dolan \& Mathieu~\cite{dm99}); {\it ii.} An initial microturbulence 
$\xi$=1.5~km/sec was adopted for all stars; {\it iii.}
Finally, surface gravities for the ONC stars and Dolan~24
were estimated from the relation
between M, L, and \teff~
($\log$~g=4.44+$\log(\rm M/\rm M_\odot)-\log(\rm L/\rm L_\odot)+
4\log \rm T_{\rm eff}-15.0447$). For the ONC stars
masses and luminosities were
taken from Getman et al.~(\cite{get05}); for the star Dolan 24
in $\lambda$ Ori we derived the 
bolometric luminosity from the I$_{\rm C}$ magnitude, 
applying the bolometric correction by Bessell 
(1991) and adopting a distance to the cluster equal to 400~pc.
Mass was then estimated using the PMS evolutionary tracks 
by Baraffe et al.~(\cite{bar98}) with $\alpha=1$. 
For HD~294297 we adopted the $\log$~g value of Cunha et al.
(\cite{cun98}).

As usually done, final temperatures were derived by
removing the trends between abundance values ($\log$~n(Fe))
and excitation potential ($\chi$) in MOOG.
Similarly, final $\xi$  values were derived by removing the trend between 
$\log$~n(Fe) and measured EWs. 
In both cases a 2$\sigma$ clipping was applied
to the initial line list before removing the trends. 
On the other hand, because of the lack of suitable Fe~{\sc ii} lines,
$\log$~g values could not be optimized for any of the stars and initial
values were adopted.

Final stellar parameters are listed in Tables~\ref{onc_parameters} and
\ref{sigma_parameters}. We first note that 
for HD~294297 we retrieve the same \teff~as
Cunha et al.~(\cite{cun98}). Then, the comparison of
Tables~\ref{onc_targets} and \ref{sigma_targets} with 
Tables~\ref{onc_parameters} and \ref{sigma_parameters} shows that the maximum
difference between final and initial effective temperatures is on the order
of 100~K, with the exception of star h673 ($\Delta$\teff$\sim 300$~K)
and Dolan24 ($\Delta$\teff=270~K). All final temperatures are warmer than
initial ones. 
Tables~\ref{onc_parameters} and \ref{sigma_parameters} also indicate that the
final microturbulence values are in the range $\sim 1.5-1.7$~km/s, only
slightly larger than those normally found for main sequence dwarfs of
similar temperature. 
For the PMS stars in their sample, and in particular
for the three ONC members,
Santos et al.~(\cite{santos08}) instead found 
higher values (up to $\xi=2.5$~km/sec) and a larger star-to-star scatter. 
They suggested that the discrepancy between their
microturbulence values and those of main sequence stars might be due to
the effect of strong magnetic fields that characterize very young
stars. Our results do not seem to support
this hypothesis, since the
microturbulence values for our sample stars are only slightly higher than 
that of the main sequence member of IC~2391 (see Sect.~\ref{errors} below).
Note that use of higher values of the
microturbulence would lead to lower [Fe/H] values.

The iron abundance for each line was obtained based on measured EWs
and adopted stellar parameters.
Final abundances for each star were then determined as the
mean abundance from the different lines.
\subsubsection{Spectral synthesis}
The spectral synthesis for the Fe analysis was performed in a 20~\AA~ 
wavelength range around the Li~{\sc i}
feature at 6707.78~\AA, from 6695~\AA to 6715~\AA. 
As in the case of the EWs, we first carried out an analysis
of the solar spectra acquired with UVES and Giraffe.
When necessary, we modified $\log gf$ values to get a good fit
of the solar spectrum with the standard solar abundances of
Anders \& Grevesse (\cite{ag89}).

The synthesis was carried out for all but one star
observed with UVES, in order to cross-check [Fe/H] from EWs; spectral
synthesis was also performed
for the two ONC Giraffe spectra, whose lower resolution and reduced spectral
range did not allow us to carry out the EW analysis. Also note that,
due to the relatively high rotation, [Fe/H] for h907 in ONC
could be measured only from the synthesis analysis.
Instead, for the star h223a in the ONC 
we could not apply the synthesis method because part
of the bluest order, including the wavelength range from 6695
to 6740~\AA, fell out of the CCD.

\begin{figure*}
\includegraphics[width=16cm,angle=0]{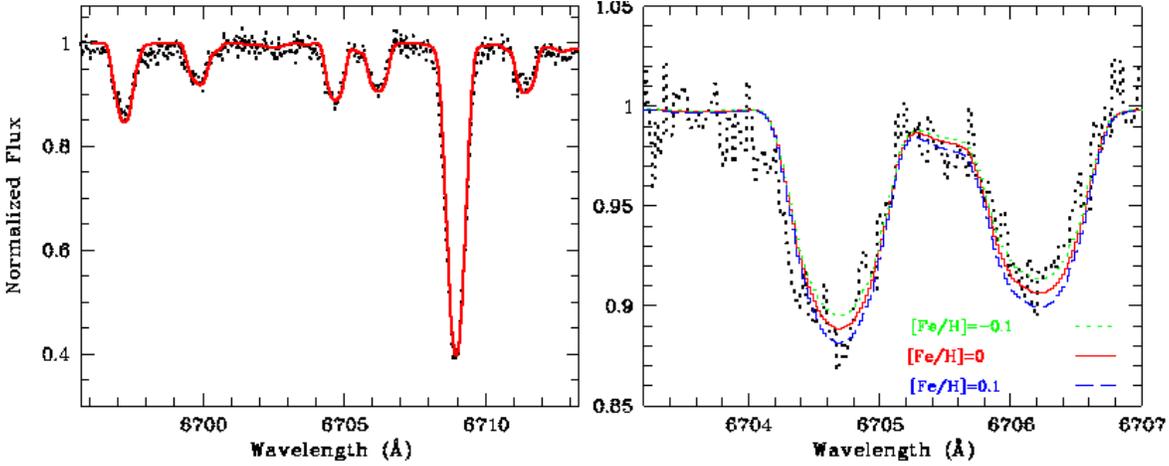}
\caption{Left panel:
Observed spectrum (dashed line) and best-fit synthetic spectrum
(solid line, [Fe/H]=0) for the star h673.
Right panel: Zoom on a restricted spectral region.
Spectral synthesis with three different values of the metallicity are shown: 
[Fe/H]=0 (solid, red), 0.1 (dashed, blue), $-$0.1 (dotted, green).}
\label{synt_uves}
\end{figure*}

\begin{figure*}
\includegraphics[width=17cm,angle=0]{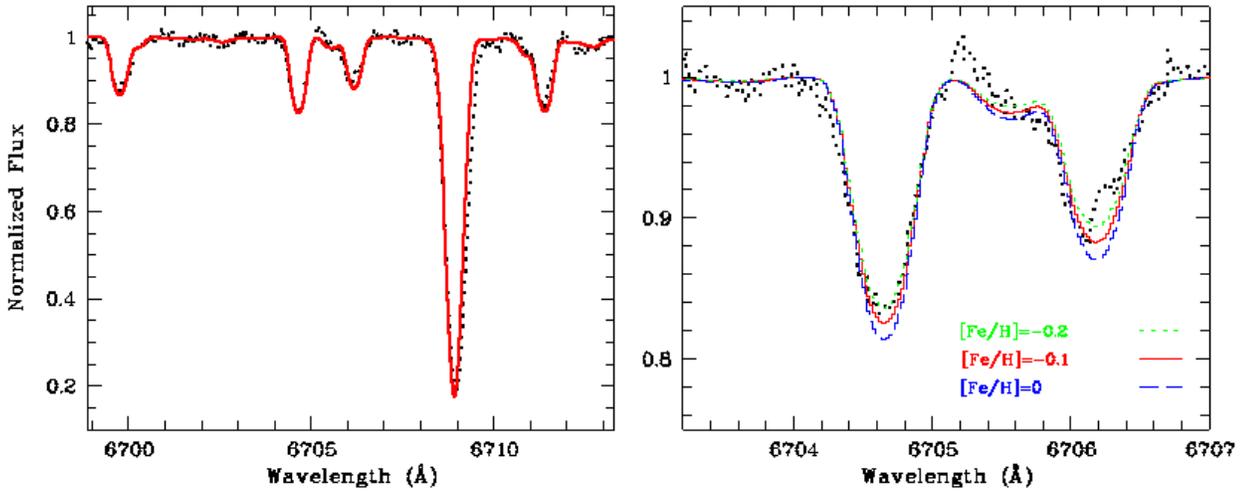}
\caption{Same as Fig.~\ref{synt_uves}, 
but the observed and synthetic spectra
of star h683 are shown. Note that the [Fe/H] from EWs for this star is slightly
larger than that from the synthesis; the figure shows that the synthetic
spectrum with [Fe/H]=$-0.1$ better fits the observed spectrum.
}
\label{synt_683}
\end{figure*}

\begin{figure*}
\includegraphics[width=17cm,angle=0]{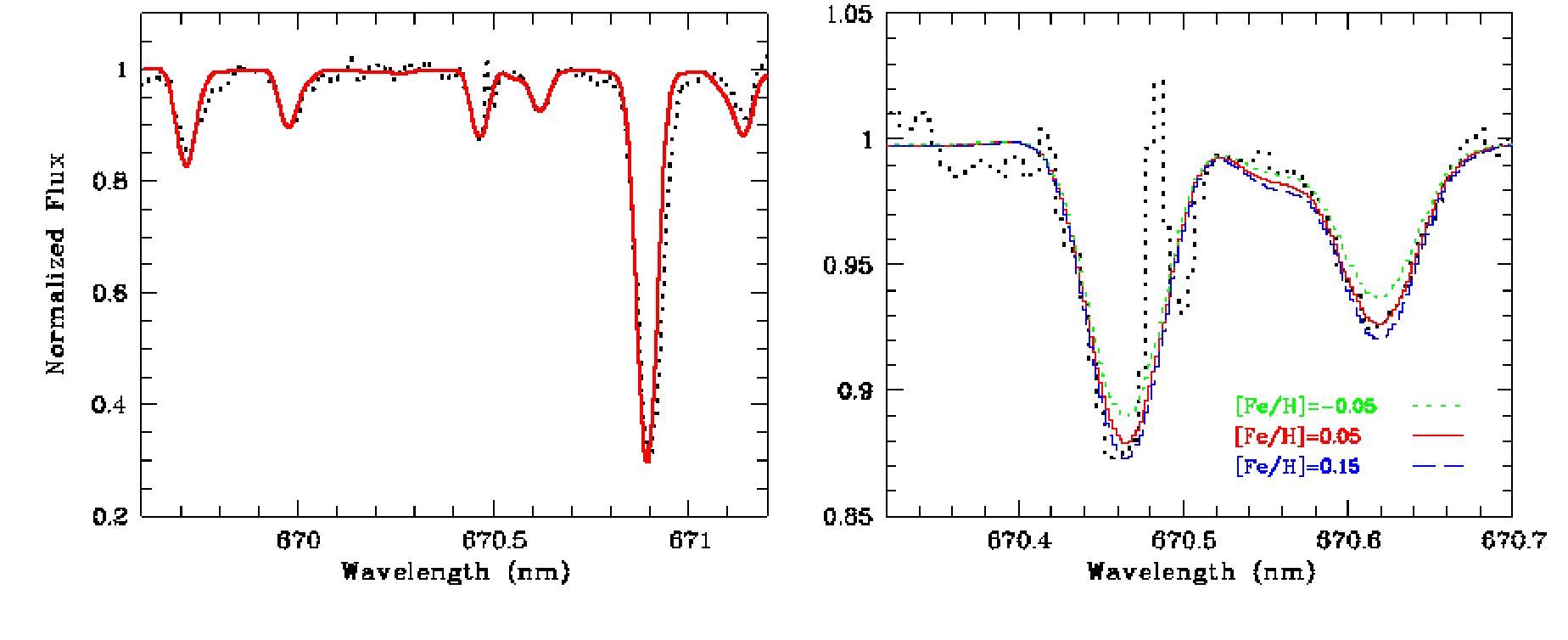}
\caption{Same as Figs.~\ref{synt_uves} and \ref{synt_683}, 
but the observed and synthetic spectra
of star h664 observed with Giraffe are shown. Note that Giraffe wavelength
scale is in nm rather than in \AA.}
\label{synt_giraffe}
\end{figure*}

For stars observed with UVES, the synthesis was carried out adopting
the final stellar parameters determined with the EW analysis;
on the other hand, stellar parameters could not be optimized for
stars observed with Giraffe and for h907, for which we did not carry
out the EW analysis. Therefore, for
these stars we adopted initial \teff~values and a microturbulence $\xi=
1.6$~km/s, which is typical of our UVES sample.
Also, for each UVES spectrum, we first computed
a synthetic spectrum with the metallicity determined through the EW 
analysis; then, if necessary, we changed the metallicity
until the best fit of the observed spectrum was obtained.
For the analysis of the Giraffe spectra and that of h907
we started with a solar metallicity.
Examples of spectral synthesis in the 6695--6715~\AA~region
for UVES and Giraffe spectra are shown
in Figs.~\ref{synt_uves} (star h673),~\ref{synt_683} (star h683),
and \ref{synt_giraffe} (star
h664), respectively. In all figures, both the whole spectral region
and a zoom around two Fe~{\sc i} lines are displayed. Those two lines
have different excitation potentials and thus their strength has
a different dependence
on effective temperature. A large error in the determination of the latter
would thus be identified in the comparison of synthetic and observed spectra.
We also note that the red side of the Li~{\sc i}~6707.8~\AA~feature
is not well fitted in the spectra of stars h683 and h664 (Figs.~\ref{synt_683}
and \ref{synt_giraffe}). Although a detailed investigation of this effect
is outside the purposes of this paper, we suggest that it might be due
to the contribution of $^6$Li to the lithium feature. $^6$Li is indeed not
included in our current line list.
We finally notice that, as a by-product,
spectral synthesis also allowed us to estimate
projected rotational velocities (v$\sin i$) for the sample stars. The latter
are listed in Column~6 of Tables~\ref{onc_parameters} and 
\ref{sigma_parameters}.
\subsection{Errors}\label{errors}
Three sources of random errors are present in the EW analysis:
(i) $\sigma_1$, the standard deviation around the mean
due to differences in abundances
that come from each line; $\sigma_1$ should be representative of errors in EWs 
and $\log$~gf values; (ii) $\sigma_2$: errors due to uncertainties on 
the stellar parameters.
We found that for all the stars variations in T$_{\rm eff}$ greater
than 100 K would produce significant trends in $\log$~n(Fe) versus 
excitation potential, while variations greater than
0.15 km/s in $\xi$ would introduce a trend between $\log$~n(Fe) and 
measured EWs; we thus assumed 100~K and 0.15 km/s as typical uncertainties
in \teff~and $\xi$. As to gravity, we adopted a typical uncertainty of 
0.25~dex. Resulting errors in [Fe/H] were then
estimated by varying one parameter at the time and quadratically adding
the resulting errors. We found that
a change in T$_{\rm eff}$ of $\pm$ 100 K results in a change of 
Fe abundance from $\pm 0.01$ to $\pm 0.04$ dex;
a change in $\xi$ of $\pm$ 0.15 km/s causes a change of 
$\pm 0.03$ dex, and a variation of 0.25 dex in $\log$~g results 
in a change from $\pm 0.04$ to $\pm 0.07$~dex in abundance;
(iii) $\sigma_3$: errors due to uncertainties on
veiling determination. As shown in Tables~\ref{onc_parameters} and
\ref{sigma_parameters}, the last
are very small and lead to an error in the final [Fe/H] on the order of
$0.01-0.02$~dex, much below the other two errors. 
Also note that those small errors on veiling do not
introduce any effect on the determination of stellar parameters.

The abundance uncertainties are related to the best fit 
determination ($\sigma1$) and to the 
uncertainties in stellar parameters ($\sigma2$),
also in the case of spectral synthesis analysis.
Assuming the same uncertainties in stellar parameters derived from the EW
analysis, we found that
a change in T$_{\rm eff}$ of $\pm 100$~K
results in a difference in Fe abundance from $\pm$0.05 to $\pm$0.1 dex, 
while a change of $\xi$ of $\pm$ 0.15~km/s causes a variation from $\pm 0.03$ 
to $\pm 0.05$ dex
in abundance values. The effect of gravity is instead negligible (below
0.02~dex).  

Systematic errors that could affect the analysis
of cool stars were estimated by analyzing with the same method
a star with similar parameters as our sample stars and known metallicity.
We chose a member of the open cluster IC~2391 (VXR76A)
whose metallicity has been
estimated with an independent method (different line list and code)
by Randich et al.~(\cite{R01} -[Fe/H]$=0.0\pm 0.09 \pm 0.09$). 
For the analysis of VXR76A we used the same spectrum
employed by Randich et al., characterized by a spectral 
resolution R=43800, very 
similar to that of the UVES spectra. We derived the metallicity of VXR76A
using both EWs and the two line lists and spectral synthesis. The results
are listed in the last raw of Table~\ref{onc_parameters}. In all cases
the agreement with the previous determination is excellent:
namely, [Fe/H]$=0.0\pm 0.06 \pm 0.08$ from the EWs and the CD3 line list,
[Fe/H]$=-0.01\pm 0.08\pm 0.08$ from the EWs and CD4 line list, and
[Fe/H]$=0\pm 0.1\pm 0.1$ from the spectral synthesis. 
Furthermore,
in a more recent analysis of a wider sample of stars and better quality
spectra of IC~2391, D'Orazi \& Randich (\cite{dr09})
not only confirmed a solar metallicity for this
cluster, but also found that
VXR76A shares the same metallicity as other, warmer members in IC~2391.
This suggests that our analysis should not be affected by major systematic 
errors and that no major offset should be present between the metallicity
of cool and warmer stars.
As for Dolan~24, the method/line list was already 
tested by Randich et al.~(\cite{randich_m67}) on M~67 members.
\section{Results and discussion}
\subsection{The metallicity of the ONC}
The results of our analysis are summarized in  Tables~\ref{onc_parameters} 
and \ref{sigma_parameters}, where we give the veiling value $r$ (Col~2), 
the adopted
T$_{\rm eff}$, $\log g$, and $\xi$ values (Cols.~3--5), v$\sin i$ (Col.~6), 
[Fe/H] derived from EWs analysis (Col.~7), along with the number of lines
employed for the analysis. [Fe/H] values from
spectral synthesis are listed in Col.~8.

First, the two tables show that, in general,
[Fe/H] values from EWs and spectral synthesis
well agree within the uncertainties.
We derived the mean ONC metallicity considering
the more secure [Fe/H] from EWs for stars with both determinations; [Fe/H]
from synthesis was instead considered
in the three cases for which the EW analysis
could not be performed. We find
[Fe/H]$_{\rm av}=-0.01 \pm 0.04$, i.e., the ONC has a
solar metallicity. Including only stars with [Fe/H] from EWs, we
would obtain [Fe/H]$_{\rm av}=-0.02\pm 0.05$.

\begin{figure*}
\includegraphics[height=16cm,angle=270]{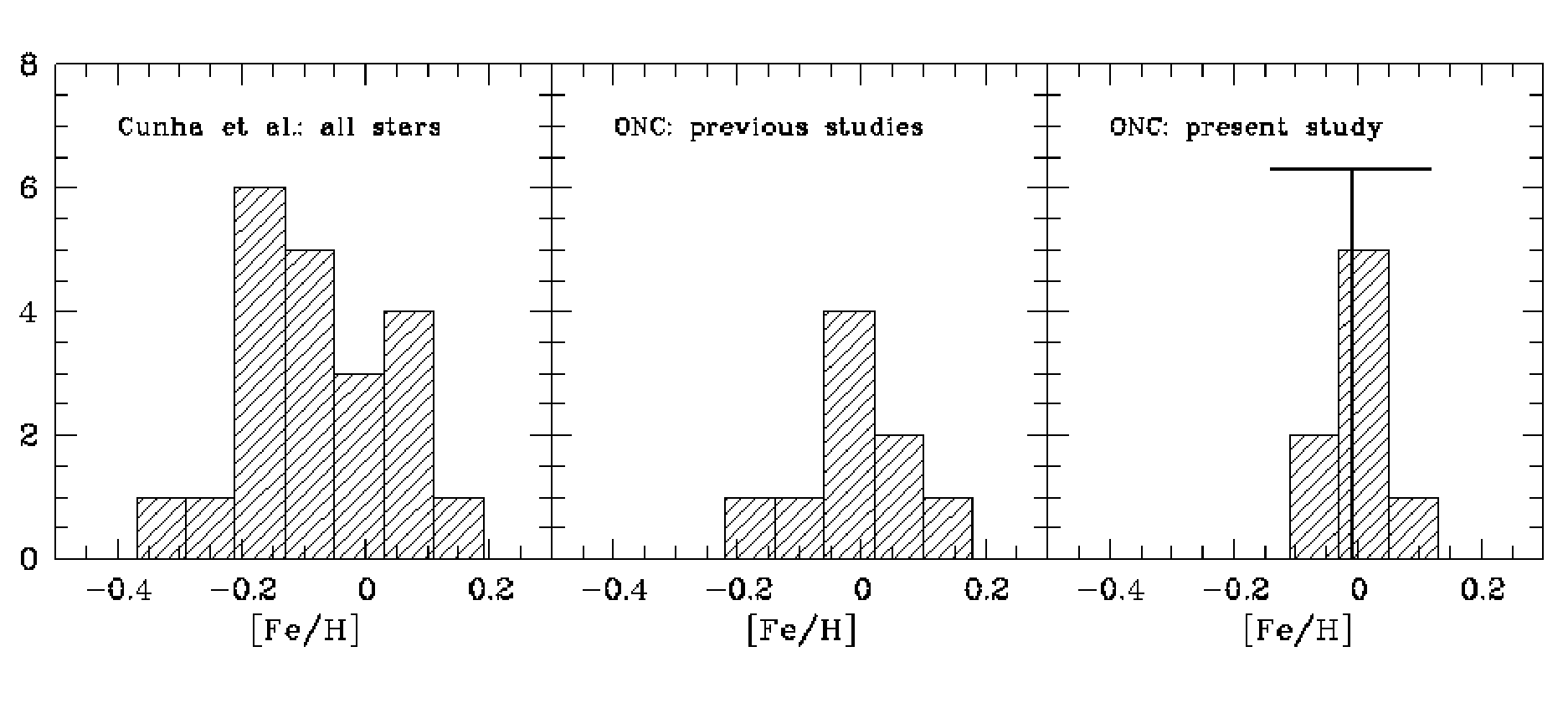}
\caption{
Left panel: [Fe/H] distribution
across all the four Orion subgroups from Cunha et al. (\cite{cun94},
\cite{cun98});
middle panel: [Fe/H] distribution of the ONC only, obtained
considering four different studies (Cunha \& Lambert~\cite{cun94};
Cunha et al. \cite{cun98}; Padgett 
\cite{padgett}; Santos et al.~\cite{santos08}); right panel: 
[Fe/H] distribution of the ONC from our study. We also indicate our
average for the ONC (vertical line) and the typical random error on
[Fe/H] of individual stars.
}\label{histo}
\end{figure*}

As mentioned, Santos et al. (\cite{santos08}), based on three stars,
derived a significantly lower average 
metallicity for the ONC ([Fe/H]$=-0.13\pm 0.06$), 
although the [Fe/H] of the most metal-rich star in their sample is comparable
to the most metal-poor one in our own sample.
The discrepancy is likely due to the higher microturbulence
values obtained in that study (see Sect.~\ref{analysis}). As in our case,
Santos et al. have used FLAMES with the fiber link to UVES and
their spectra are of similar quality. For the analysis
they have employed the same code and model atmospheres; however,
they adopted a different line list and $\sigma$-clipping criterion.
We suggest that this affects
the microturbulence determination and thus the final [Fe/H].
We stress again that
the comparison with VXR76A provides confidence in our results.
\setcounter{table}{2}
\begin{table*}
\caption{Final stellar parameters and metallicity values derived for ONC 
stars.$^1$}
\label{onc_parameters}
\begin{tabular}{ccccccrr}
\hline
star     & $r$  & T$_{\rm eff}$  &  $\log$~g & $\xi$ & v$\sin i$ & \multicolumn{2}{c}{[Fe/H] $\pm$ $\sigma_1$ $\pm$ $\sigma_2$} \\
   &         & (K) &  & (km/s) & (km/s) & (EWs) & (Spec. Synth.)\\
 (1) & (2) & (3)  & (4)  & (5)  &  (6)  & (7) & (8) \\
\hline
223a &  $0.02 \pm 0.04$ & 4450    & 3.8 & 1.6 & --- & $0.03\pm0.06\pm 0.07$ (18)& ---\\
268  &  $0.03 \pm 0.04$ & 4300    & 3.9 & 1.6 & 10$\pm3 $ & $-0.08\pm 0.16\pm 0.09$ (25)& $-0.1\pm 0.1\pm 0.1$\\
278  &  $0.8  \pm 0.07$ & 4395    & --- & --- &--- & --- & ---\\
487  &  $0.02 \pm 0.02$ & 4300    & 3.9 & 1.7 & 15$\pm3 $ & $-0.07\pm 0.08\pm 0.09$ (18) & $0\pm 0.1\pm 0.09$\\
673  &  $ 0.04 \pm 0.03$ & 4700   & 4.0 & 1.5$^*$ & 20$\pm3 $ & $0.0\pm 0.1\pm 0.07$ (18)& $0\pm 0.1 \pm 0.08$\\
683  &  $ 0.0 \pm 0.02$ & 4250    & 3.3 & 1.6 & 13$\pm3 $ & $0.01\pm 0.1\pm 0.08$ (22) & $-0.1\pm 0.1 \pm 0.09$\\
907  &  $0.01 \pm 0.03$ & 4775    & 4.1 & 1.5 & 35$\pm3 $ & --- & $0.0\pm 0.1 \pm 0.11$\\
     &       &     &  &  & & & \\
664  &  0$^{**}$ &     4200	 & 4.1 & 1.6 & 13$\pm3 $ & --- & $0.05\pm 0.1\pm 0.1$\\
1020 & 0$^{**}$ &      4200	& 3.8 & 1.6 & 13$\pm3 $ & --- & $0.0\pm 0.1\pm 0.1$\\
     &     &      		&   &  & & & \\
VXR76A (list$_{\rm CD3}$)& --- &      4340 & 4.5  & 1.2  & 8$\pm$3 & $0.00\pm 0.06 \pm 0.09$ (25) & $0 \pm 0.1\pm 0.1$\\
VXR76A (list$_{\rm CD4}$)& --- &      4400	 & 4.5  & 1.3 & 8$\pm$3 & $-0.01\pm 0.08\pm 0.08$ (23)& $0\pm 0.1\pm 0.1$\\
     &     &      		&   &  & & & \\
\multicolumn{8}{l}{${^*}$The initial microturbulence was adopted
for this star, since $\xi$ could not be optimized due to the lack of 
}\\
\multicolumn{8}{l}{
lines covering a wide enough EW interval.}\\
\multicolumn{8}{l}{${^{**}}r$ values for stars 664 and 1020 come from Palla
et al.~(\cite{pal05}).}\\
\hline	     
\end{tabular}\\
\footnotemark{In the table we list
the star ID, the veiling $r$, adopted stellar parameters,
projected rotational velocities, and the final
[Fe/H] values, from the EW and synthesis analysis,
along with $\sigma_1$ and $\sigma_2$ errors (see text).
For [Fe/H] from EWs we also provide in parenthesis the 
number of employed Fe~I lines.}
\end{table*}
\setcounter{table}{3}
\begin{table*}
\caption{Same as Table~\ref{onc_parameters}, but the results
for $\lambda$~Ori targets and star HD~294297 are listed.
}\label{sigma_parameters}
\begin{tabular}{ccccccrr}
\hline
star     & $r$  & T$_{\rm eff}$  &  $\log$~g & $\xi$ & v$\sin i$ & \multicolumn{2}{c}{[Fe/H] $\pm$ $\sigma_1$ $\pm$ $\sigma_2$} \\
         &    & (K)   &  & (km/s) & (km/s) & (EWs) & (Spec. Synth.)\\
 (1) & (2) & (3)  & (4)  & (5)  &  (6)  & (7) & (8) \\
      &  &         &         &           &           &          &    \\  \hline
HD~294297  & --  & 6100 & 4.0 & 1.4 & $<7.5$ & $-0.16\pm 0.03\pm 0.07$ (34)& $-0.15\pm 0.05\pm 0.08$\\
             &           &         &           &    &       &          &    \\ 
Dolan24  &  0.05$\pm$0.03  & 5350 & 4.3 & 1.4 & $<$7.5 & $0.01\pm 0.02\pm 0.06$ (54)& $0.1\pm 0.1\pm 0.08$\\
\hline	     
\end{tabular}
\end{table*}
As discussed in Sect.~1, the metallicity of the ONC has so far been
poorly constrained, with some studies suggesting an underabundance
with respect to the Sun and others finding solar or even over-solar
[Fe/H] values. Iron has been determined in eight low-mass stars 
(1 from Cunha et al.~\cite{cun98}, four from Padgett~\cite{padgett},
and three from Santos et al.~\cite{santos08}) and
one B-type star (Cunha \& Lambert~\cite{cun94}),
but the [Fe/H] values are not homogeneous,
since they were measured with different techniques.
Our study, not only almost
doubles the number of ONC low-mass members with a metallicity
measurement, but our [Fe/H] are homogeneous,
thus allowing us 
to secure the average metallicity of the ONC on more solid
grounds. The left panel of
Fig.~\ref{histo} shows the distribution of [Fe/H] for the 
sample of Cunha et al. (\cite{cun94}, \cite{cun98}) including
both early- and late-type stars in all the four subgroups; in the middle
panel we plot the [Fe/H] distribution of ONC only, considering all previous
determinations from
the literature; finally, the distribution of our own measurements for
ONC members is displayed in the right-hand panel, along with our average
and the typical error on [Fe/H] of individual stars.
The figure shows that the distribution is very broad when including
other subgroups besides the ONC, possibly implying a group-to-group
difference; the distribution becomes narrower when considering
the ONC only and measurements from different groups, 
but still spans $\sim 0.4$~dex in [Fe/H].
The distribution of our own [Fe/H] values is instead much narrower
and the small dispersion still present is completely
due to measurement uncertainties.

As to OB1b, our [Fe/H] determination for
star HD~294297 is in excellent agreement with the value
of Cunha et al. (\cite{cun98} -[Fe/H]$=-0.19$). For the same star
Gonz\'alez-Hernand\'ez et al.~(\cite{gonzalez}) 
derived [Fe/H]$=-0.09\pm 0.1$ (star
SO000041 in their list), also in good agreement. All the three measurements
indicate a sub-solar metallicity, much
below our average for the ONC.
In our study the [Fe/H] values for this star and the ONC are
on the same scale; we mention in passing that, by using for the analysis
of HD~294297 the same subset of lines used for the ONC, we would still obtain
for this star the same [Fe/H]. This suggests that the difference
between the ONC stars and HD~294297 is real and that the
OB1b group might
be more metal poor than the ONC. We further develop this point in the next
section.

Finally, the metallicity of $\lambda$~Ori is likely
solar; whereas this represents the first measurement
of the metallicity in this cluster, it is
based on one star only and, obviously, additional measurements are needed.
\subsection{The iron distribution in Orion}
\begin{figure*}
\includegraphics[width=17cm]{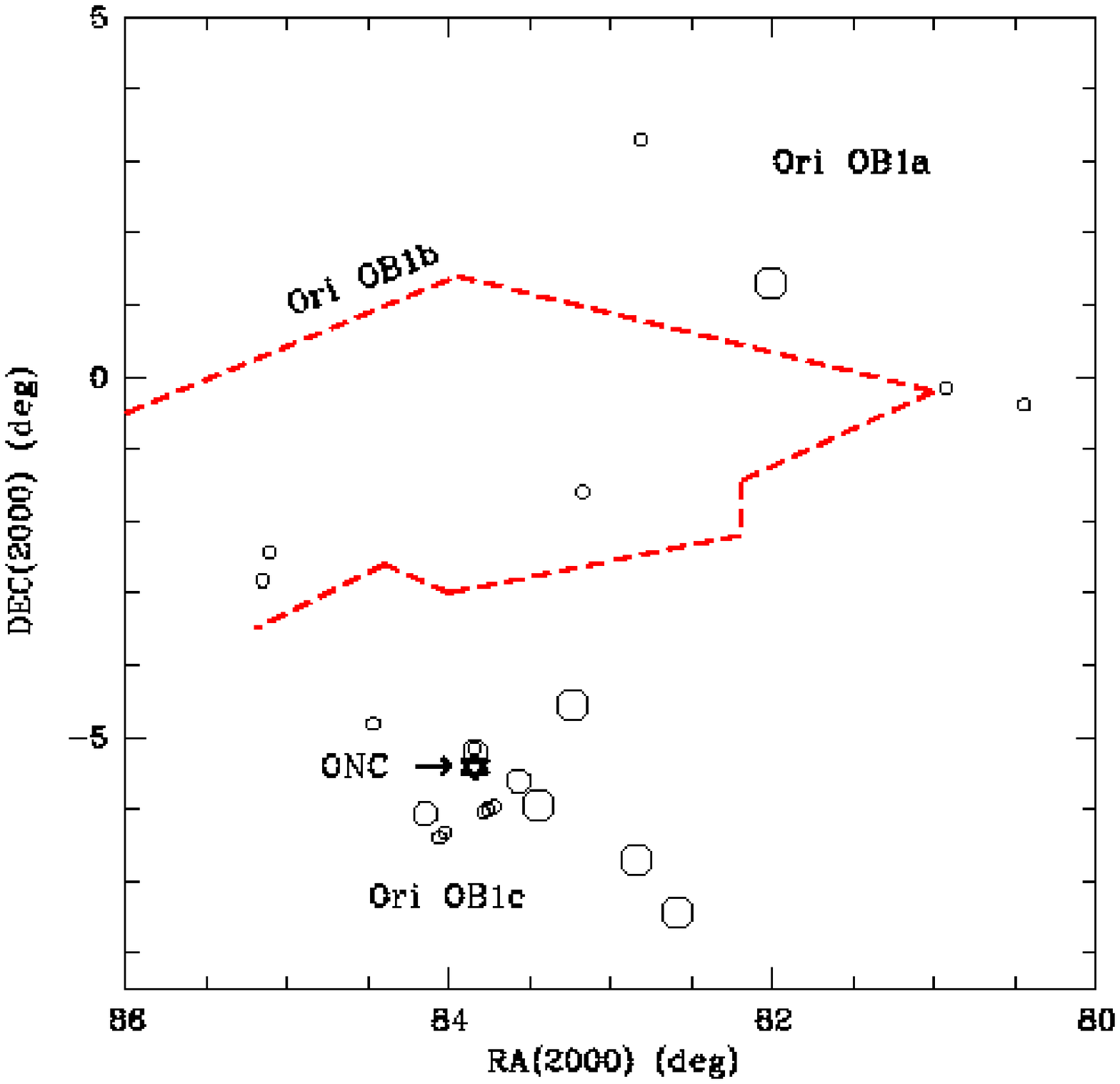}
\caption{Spatial distribution of members 
of the Orion association
with [Fe/H] measurements. Open symbols indicate members of the OB1a,1b, 1c
subgroups, while our average for the ONC is represented by the star symbol.
Symbols with different size denote three different metallicity bins:
[Fe/H]$\leq -0.05$, $-0.05 <$[Fe/H]$\leq 0.03$, and [Fe/H]$>0.03$, from
the smallest to the largest bin. 
The central bin corresponds to the average metallicity of the 
ONC $\pm 1\sigma$.
The dashed line outlines the boundaries of the Ori OB1b subgroup
following Warren \& Hesser (\cite{wh77}).
}\label{distribution}
\end{figure*}

In Fig.~\ref{distribution} we plot the spatial distribution of 
stars in the different Orion subgroups with an
available metallicity determination; stars
in three metallicity bins are indicated with symbols of different size.
For the ONC we show our average measurement.

The figure indicates that the three stars
belonging to the OB1b subgroup all
have [Fe/H] below the average of the ONC and,
possibly, of the 1c subgroup, whose metallicity is characterized
by a larger uncertainty and dispersion. 
Two of the stars in the OB1b sample are B-type
stars from Cunha \& Lambert (\cite{cun94}), while the third one
is HD~294297; for this star we adopt our own [Fe/H] determination. 
The resulting average
metallicity for the 1b subgroup is
[Fe/H]$=-0.10\pm 0.05$, 2$\sigma$ below our determination for the ONC.
The difference is not big and based on very few objects only, two of
which are early-type stars.
Nevertheless, it suggests that the Orion region may not be
characterized by a single value of [Fe/H].
As well known, Fe is mostly produced by type-I Supernovae (SNe), 
whose lifetimes
are much greater than the age of the
Orion association ($\sim 10^7$~years). On the other hand,
massive, short-lived stars exploding as type-II SNe might pollute
a molecular cloud close to OB associations. Although SNe II are mainly
producer of $\alpha$~elements like O, Mg, and Si,
a small amount of Fe is also produced. As mentioned in Sect.~\ref{status},
Cunha \& Lambert (\cite{cun92}) and Cunha et al.~(\cite{cun98})
reported evidence of star-to-star variations in O and Si 
abundances within Orion,
with a few O/Si-enhanced stars more
centrally located in the Trapezium region than the widespread low
O/Si ones; they suggested a simple scenario
of supernova self-enrichment across the Orion association over the
past 10 Myr. With the caveat that the Fe yields from SNe II are
highly uncertain, they estimated that the enhancement in
oxygen of $\sim$~0.3~dex would correspond to
an enhancement in Fe of $\sim 0.06$~dex. Whereas in their study
they did not find such a variation,
the $\sim 0.1$~dex difference in iron
that we measure between the ONC and 1b subgroup is consistent
with this the scenario.
The OB1a subgroup might also be metal-poor;
three of the four members (all from Cunha \& Lambert~\cite{cun94})
have [Fe/H] more than $1 \sigma$~below the ONC, while one star
is overabundant with respect to the Sun and the ONC.

Finally, Fig.~\ref {distribution} shows that a dispersion in metallicity
is present in OB1c. We suggest that 
the dispersion is likely due to both
systematic differences between different studies (i.e., Cunha et al. versus
Padgett) and to large uncertainties within a given study, 
as we have shown to be the case for the ONC.
\subsection{The lack of metal-rich stars}
One of the goals of our project is the search for metal-rich young
low-mass stars. The first results presented here indicate that the
ONC has a solar metallicity;
the dispersion we find within the cluster is consistent with
measurement errors and
none of the stars in our sample is metal-rich. Under the assumption
that our analysis is not affected by major systematic errors,
as the comparison with IC~2391 confirms,
two opposite hypothesis can be suggested to explain
the lack of such a population.
The simplest and most straightforward
one is that they do not exist and, as expected within a cluster,
the ONC is characterized by a homogeneous composition of all its 
members.
Alternatively, the [Fe/H] distribution of the ONC is inhomogeneous and
metal-rich members exist, but low number statistics prevented us
from detecting them. To this respect,
we recall that one star from Padgett (\cite{padgett}) 
has a metallicity significantly
above solar ([Fe/H]=$0.14 \pm 0.18$) and above our mean for the ONC.
The final answer on the possible presence of metal-rich stars can only be 
provided by a metallicity determination in a much wider sample
of stars, including also a re-analysis of
the possible metal-rich ONC member from the study of Padgett. 
We caveat however that a large metallicity dispersion within
the ONC itself would be difficult to explain without assuming ad-hoc processes.
To our knowledge, abundance patterns in all open clusters
so far investigated
are rather homogeneous and no metal-rich members of solar-metallicity
clusters or, viceversa, solar-metallicity members of metal-rich clusters
have been detected.

\begin{figure*}
\includegraphics[width=8.5cm, angle=-90]{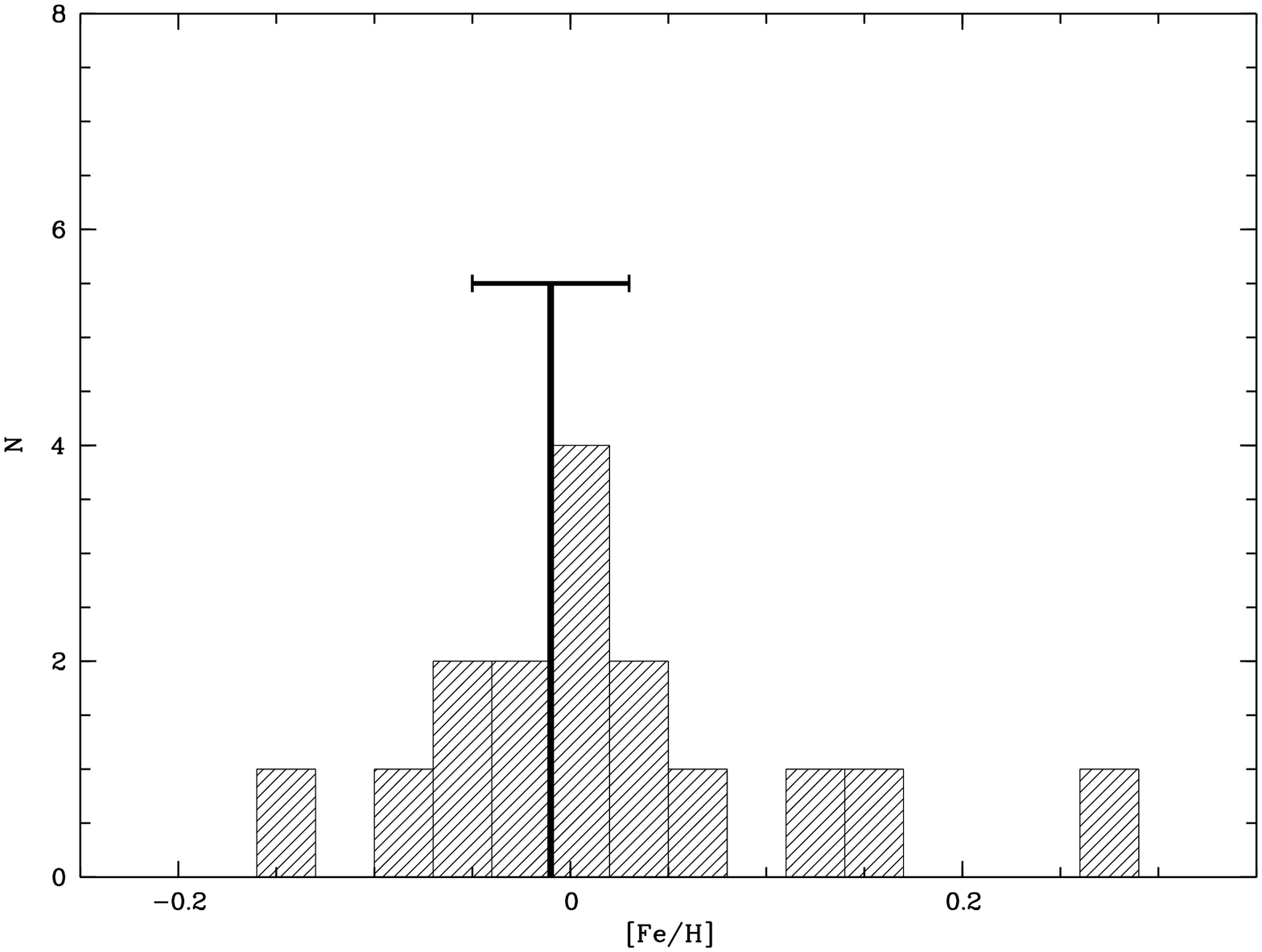}
\caption{Metallicity distribution of clusters within 500 pc
from the Sun with a secure measurement of the metallicity; the average
$\pm 1 \sigma$ for the ONC is indicated as a vertical bar.}
\label{fig_clusters}
\end{figure*}

On more general grounds, the question arises on the existence of
metal-rich young associations and SFRs. None has been
detected so-far (see the discussion in Santos et al.~\cite{santos08}). 
In Fig.~\ref{fig_clusters} we show the metallicity distribution 
of open
clusters within 500~pc from the Sun with an available spectroscopic [Fe/H]
determination. 
The 16 clusters included in the sample have ages in the interval 
30~Myr--2~Gyr; the metallicity for all but one has been
derived from high-resolution spectra of 
low-mass stars and typical errors on the average [Fe/H] values
are in the range 0.02--0.1~dex. The figure shows that
the distribution peaks at solar [Fe/H], with
a not negligible tail at high values; specifically,
metallicities vary between [Fe/H]$\sim -0.15$ 
and 0.4 with an average value [Fe/H]~$=0.02\pm 0.1$. 
In other words, the majority of the clusters in the solar neighborhood
share the same metallicity as the Sun and
the ONC is not peculiar.
On the other hand, metal--rich open clusters do exist
in the solar circle, and their number, and thus probability
of finding them, is in principle not negligible (3/16 or 19 \%
of the clusters have [Fe/H]$> 0.1$).
However, whereas a strict [Fe/H] versus age relationship 
does not hold, 
the four clusters with [Fe/H]$\geq 0.05$ are all older
than 200~Myr; furthermore, two of them (Hyades and Praesepe) belong to
the Hyades super-cluster or stream (Eggen~\cite{eggen}; 
Fam\ae y et al. \cite{fama07}) and might have a peculiar origin. 
The majority of the clusters with close-to-solar metallicity are instead
part of the Local Association (or Pleiades moving group)
or of the IC~2391 supercluster, to which various young
associations such as Scorpius-Centaurus are also associated. 
Finally, the
open clusters IC~4665 and IC~2391, both with a solar iron content
 are likely members of the
Gould Belt (Piskunov et al.~\cite{pisk06}). 
In other words, both kinematic and age considerations suggest that it might
not be easy to find a metal-rich young association in the solar
neighborhood. 
\section{Concluding remarks}
Using FLAMES/UVES and Giraffe spectra, 
we determined iron abundances in eight low-mass members of the ONC,
one star in the $\lambda$~Ori cluster, and one member of
the Orion OB1b subgroup.
The average metallicity of the ONC is
[Fe/H]$=-0.01\pm 0.04$; the metallicity of Dolan~24 in $\lambda$~Ori
is also 
solar, while the star belonging to the OB1b subgroup is significantly
more metal-poor than the ONC. 
The $\sim 0.1$~dex difference in the
Fe content of the ONC and the OB1b subgroup is consistent with the
self-pollution scenario originally proposed by Cunha and collaborators
(\cite{cun92},~\cite{cun94},~\cite{cun98}) for the Orion region. 

None of our sample stars nor any of the SFRs so far studied is metal-rich.
At the same time, known metal-rich open clusters are all older than 200 Myr, 
and none is spatially or kinematically
associated to young nearby associations and SFRs.
This suggests that finding
the metal-rich SFRs among the best studied ones in the
solar neighborhood might not be easy or doomed to failure.
We suggest that
in order to find them, one should possibly look
outside the Gould Belt and the Local Association. High-spectral
resolution observations of low-mass stars in SFRs at distances larger than 
500~pc might indeed be feasible with current instrumentation.
\begin{acknowledgements}
It is a pleasure to thank N. Santos and C. Melo for useful discussions. 
We thank the referee, Luca Pasquini, for very useful comments.
This work has made extensive use of the services of WEBDA, ADS, CDS and
of the catalog by W. Dias (http://www.astro.iag.usp.br/~wilton/).
We acknowledge partial support from PRIN-INAF ``Stellar clusters: test for
stellar formation and evolution" (PI: F. Palla).
\end{acknowledgements}
\setcounter{table}{4}
\begin{longtable}{lcccrc}
\caption{Line list of 42 Fe~{\sc i} lines used for the analysis
of the ONC stars along with excitation potential ($\chi$), 
$\log$~gf values,
equivalent widths measured on the solar spectrum acquired with UVES
and corresponding Fe abundances.
Lines labeled with a `w' or `c' have been used for the warm star
HD~294297 or for the cool ONC stars only.
}\label{sun}\\
\hline
$\lambda$ & $\chi$ & $\log$~gf & EW$_{\odot}$ & $\log$n(Fe)$_{\odot}$\\ 
(\AA)     & (EV)   &          & (m\AA) & \\ \hline
\endfirsthead
\caption{continued.}\\
\hline
$\lambda$ & $\chi$ & $\log$~gf & EW$_{\odot}$ & $\log$n(Fe)$_{\odot}$ \\ 
(\AA)     & (EV)   &          & (m\AA) & \\ \hline
\endhead
\hline\endfoot
 6703.576  &   2.760 & -3.100 &   37 & 7.53\\  
 6713.745$^{\rm w}$  &   4.790 & -1.410 &   23 & 7.50\\
 6726.673$^{\rm w}$  &   4.610 & -1.050 &   48 & 7.50\\
 6750.164  &   2.420 & -2.655 &   75 & 7.51\\
 6786.860  &   4.190 & -1.900 &   26 & 7.49\\
 6806.856  &   2.730 & -3.140 &   36 & 7.51\\
 6810.267  &   4.610 & -1.000 &   50 & 7.49 \\
 6820.374$^{\rm w}$  &   4.640 & -1.160 &   41 & 7.51\\
 6839.835  &   2.560 & -3.450 &   29 & 7.50 \\ 
 6843.655$^{\rm w}$  &   4.550 & -0.860 &   63 & 7.53\\
 6858.155  &   4.610 & -0.950 &  52 & 7.48\\
 6862.496$^{\rm w}$  &   4.560 & -1.430 &  33 &7.53\\
 6945.200  &   2.420 & -2.460 &  81 & 7.41\\
 6951.250  &   4.560 & -1.050 &  52 & 7.52\\ 
 6978.860$^{\rm w}$  &   2.480 & -2.490 &  84 &7.56\\
 6988.530  &   2.400 & -3.420 &  39 & 7.51\\
 7022.960  &   4.190 & -1.110 &  67 & 7.51\\ 
 7024.070$^{\rm w}$  &   4.070 & -1.940 &  30 &7.49\\
 7038.230$^{\rm w}$  &   4.220 & -1.130 &  62 &7.47\\  
 7083.400$^{\rm w}$  &   4.910 & -1.260 &  24 & 7.48\\ 
 7142.520$^{\rm w}$  &   4.950 & -0.930 &  39 &7.51\\  
 7219.690  &   4.070 & -1.570 &  50 & 7.52\\ 
 7221.210  &   4.560 & -1.220 &  42 & 7.49\\
 7228.700$^{\rm c}$  &   2.760 & -3.270 &  27 & 7.45\\ 
 7284.840  &   4.140 & -1.630 &  43 & 7.51\\  
 7306.570  &   4.180 & -1.550 &  43 & 7.47\\
 7401.690$^{\rm w}$  &   4.190 & -1.600 &  44 &7.54\\
 7418.670  &  4.140  & -1.440  &  51 &7.47\\  
 7421.560$^{\rm w}$  &  4.640  & -1.690  &  20 & 7.52\\  
 7447.400  &  4.950  & -0.950  &  35 & 7.44\\
 7461.530$^{\rm c}$  &  2.560  & -3.450  &  30 & 7.49\\ 
 7491.660  &  4.300  & -1.010  &  70 & 7.54\\
 7507.270  &  4.410  & -1.030  &  62 & 7.52\\ 
 7531.150  &  4.370  & -0.640  &  90 & 7.58 \\
 7547.900$^{\rm w}$  &  5.100  & -1.110  &  25 & 7.51 \\ 
 7568.910  &  4.280  & -0.900  &  76 & 7.52 \\  
 7583.800  &  3.020  & -1.930  &  85 & 7.49 \\
 7710.363  &  4.220  & -1.112  &  66 & 7.48 \\
 7751.110$^{\rm w}$  &  4.990  & -0.740 &   48 &7.50\\
 7807.910$^{\rm w}$  &  4.990  & -0.510 &   61 & 7.50\\
 7912.870  &  0.860  & -4.850 &   49 & 7.50\\
 7959.150$^{\rm w}$  &  5.030  & -1.180 &   24 &7.47\\
\hline 
\end{longtable}
{}
\end{document}